\documentclass[amsmath,amssymb,superscriptaddress,longbibliography,prb,twocolumn,floatfix]{revtex4-2}

\usepackage{amsmath,amssymb}
\usepackage[normalem]{ulem}
\usepackage{graphicx}
\usepackage{xcolor}
\usepackage{hyperref}
\hypersetup{
    colorlinks=true,
    urlcolor= blue,
    citecolor=blue,
    linkcolor= blue}

\definecolor{lightgray}{gray}{0.6}
\newif\ifptitle
\newif\ifpnumber
\newcounter{para}
\newcommand\ptitle[1]{\par\refstepcounter{para}
{\ifpnumber{\noindent\textcolor{lightgray}{\textbf{\thepara}}\indent}\fi}
{\ifptitle{\textbf{[{#1}]}}\fi}}
\pnumbertrue  

\newcommand{\Qlat}{\mathbf{Q}_{\mathrm{lat}}}
\newcommand{\Qa}{\mathbf{Q}_{\mathrm{a}}}
\newcommand{\Qb}{\mathbf{Q}_{\mathrm{b}}}
\newcommand{\Qta}{\mathbf{Q}_{\mathrm{2a}}}
\newcommand{\Vs}{V_\mathrm{s}}
\newcommand{\Is}{I_\mathrm{s}}
\newcommand{\Vrms}{V_\mathrm{rms}}
\newcommand{\EF}{E_F}

\newcommand{\pppp}{(++++)}
\newcommand{\pppm}{(+++$-$)}
\newcommand{\ppmm}{(++$--$)}
\newcommand{\pmpm}{(+$-$+$-$)}

\newcommand{\hphys}{Department of Physics, Harvard University, Cambridge, Massachusetts 02138, USA}
\newcommand{\heng}{School of Engineering \& Applied Sciences, Harvard University, Cambridge, Massachusetts 02138, USA}
\newcommand{\ames}{Ames Laboratory, Iowa State University, Ames, Iowa 50011, USA}
\newcommand{\iowa}{Department of Physics and Astronomy, Iowa State University, Ames, Iowa 50011, USA}
\newcommand{\lanl}{Theoretical Division, Los Alamos National Laboratory, Los Alamos, New Mexico 87545, USA}
\newcommand{\cin}{Center for Integrated Nanotechnology, Los Alamos National Laboratory, Los Alamos, New Mexico 87545, USA}

\begin{document}

\title{Spin-polarized imaging of strongly interacting fermions\\
in the ferrimagnetic state of Weyl candidate CeBi}

\author{Christian E. Matt}
\affiliation{\hphys}
\author{Yu Liu}
\affiliation{\hphys}
\author{Harris Pirie}
\affiliation{\hphys}
\author{Nathan C. Drucker}
\affiliation{\heng}
\author{Na Hyun Jo}
\affiliation{\ames}
\affiliation{\iowa}
\author{Brinda Kuthanazhi}
\affiliation{\ames}
\affiliation{\iowa}
\author{Zhao Huang}
\affiliation{\lanl}
\author{Christopher Lane}
\affiliation{\lanl}
\affiliation{\cin}
\author{Jian-Xin Zhu}
\affiliation{\lanl}
\affiliation{\cin}
\author{Paul C. Canfield}
\affiliation{\ames}
\affiliation{\iowa}
\author{Jennifer E. Hoffman}
\email[]{jhoffman@physics.harvard.edu}
\affiliation{\hphys}
\affiliation{\heng}

\date{\today}

\begin{abstract}
CeBi has an intricate magnetic phase diagram whose fully-polarized state has recently been suggested as a Weyl semimetal, though the role of $f$ states in promoting strong interactions has remained elusive. Here we focus on the less-studied, but also time-reversal symmetry-breaking ferrimagnetic phase of CeBi, where our density functional theory (DFT) calculations predict additional Weyl nodes near the Fermi level $\EF$. We use spin-polarized scanning tunneling microscopy and spectroscopy to image the surface ferrimagnetic order on the itinerant Bi $p$ states, indicating their orbital hybridization with localized Ce $f$ states. We observe suppression of this spin-polarized signature at $\EF$, coincident with a Fano line shape in the conductance spectra, suggesting the Bi $p$ states partially Kondo screen the $f$ magnetic moments, and this $p-f$ hybridization causes strong Fermi-level band renormalization. The $p$ band flattening is supported by our quasiparticle interference (QPI) measurements, which also show band splitting in agreement with DFT, painting a consistent picture of a strongly interacting magnetic Weyl semimetal.
\end{abstract}

\maketitle

\section{Introduction} 

\ptitle{Strongly correlated topology}
Merging strong electron interactions with topology is a new frontier for fundamental research and advanced technology~\cite{MaciejkoNatPhys2015, TokuraNatPhys2017, RachelRPP2018}. Kondo lattice systems are a promising platform for strongly correlated topological phenomena, exemplified by the recent observation of strongly-renormalized Dirac surface states in the Kondo insulator SmB$_6$~\cite{DzeroPRL2010,PirieNatPhys2020}, and the proposal for a Weyl-Kondo semimetal phase~\cite{LaiPNAS2018}.
In general, a Weyl semimetal~\cite{ArmitageRMP2018} arises when a bulk Dirac point is split into two Weyl nodes by breaking inversion or time-reversal symmetry (TRS). However, a crystal structure that breaks inversion symmetry is typically not tunable, while an applied magnetic field $B$ that breaks TRS yields only a small Zeeman energy of $\sim1$ Kelvin/Tesla for a typical $g$-factor of 2.  Materials with intrinsic magnetic order may have larger energy scales that drive the Weyl nodes farther apart and protect their well-defined chirality~\cite{ArnoldNatCom2016}. Such TRS-breaking Weyl semimetals were recently discovered in ferromagnets~\cite{MoraliScience2019,LiuScience2019,BelopolskiScience2019} and antiferromagnets~\cite{KurodaNatMat2017}. The ultimate goal is to \textit{combine} the higher energy scales and strong correlations with the practicality of external tunability~\cite{SmejkalNatPhys2018}. This goal motivates the search for topological phases in Kondo lattice compounds, which often host large spin-orbit coupling, strongly interacting electrons, and proximate field-tunable magnetic order~\cite{DoniachPhysicaB1977,SiScience2010,Coleman2015}.

\begin{figure}[ht]
    \includegraphics[width=\columnwidth]{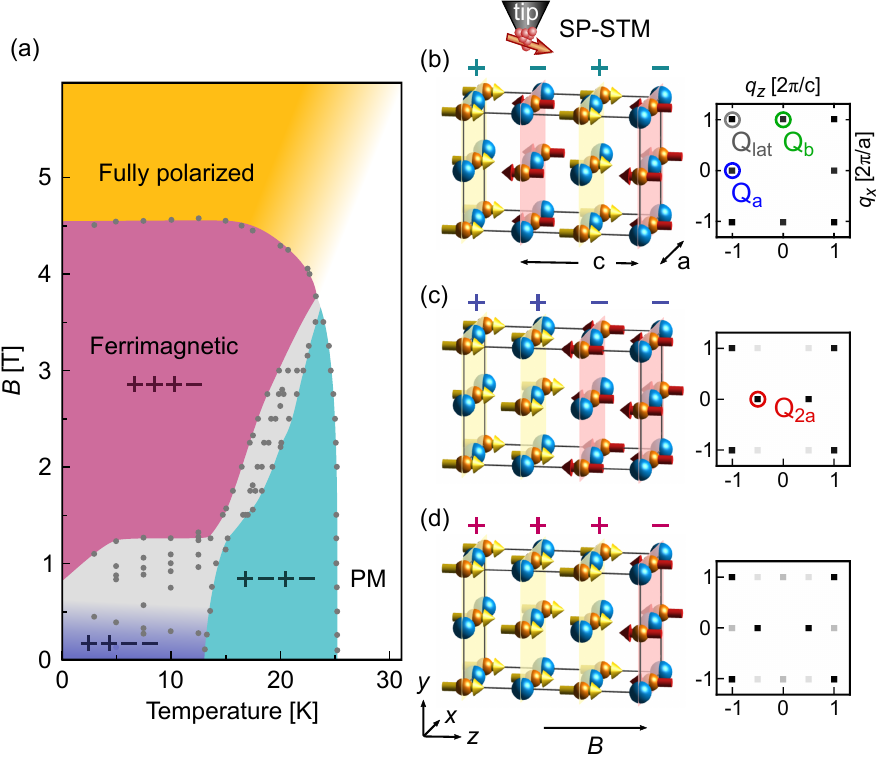}
    \caption{(a) Magnetic phase diagram of bulk CeBi, with dots from magneto-transport measurements~\cite{Kuthanazhi2021} marking an intricate cascade of transitions between magnetic orders.
    (b-d) Real-space structure and simulated Fourier transforms of the $x-z$ plane of CeBi in the \pmpm, \ppmm, and \pppm\ magnetic phases. Here, ``$+$''and ``$-$'' indicate the direction of the Ce $f$ net magnetic moments, which are ferromagnetically aligned in each $x-y$ plane with varying order along the $z$ direction. $\Qlat$ indicates the wavevector of the Ce (or Bi) sublattice, which would appear in spin-averaged STM images. Spin-polarized STM would be sensitive to additional magnetic Bragg peaks. (b) Antiferromagnetic \pmpm\ phase for $T\lesssim25$ K. (c) Antiferromagnetic \ppmm\ phase for $T\lesssim12.5$ K. (d) Ferrimagnetic \pppm\ phase with magnetic field $B$ applied along $z$.}
     \label{fig:phase_diagram}
\end{figure}

\ptitle{CeBi open questions} 
Cerium-monopnictides (Ce$X$, $X$ = As, Sb, Bi) are correlated low-carrier-density Kondo lattice systems~\cite{SuzukiPBCM1993} with cascades of magnetic phase transitions, as shown for CeBi in Fig.~\ref{fig:phase_diagram}~\cite{BartholinJPC1979, Kuthanazhi2021, HulligerJLTP1975}. Though bulk magnetic phase diagrams have been measured by neutron scattering, surface magnetic order has not been studied. Meanwhile, non-trivial band topology was predicted in CeSb~\cite{FangPRB2020, GuoNPJQM2017} and CeBi~\cite{KurodaPRL2018, HuangPRB2020}, and signatures of Weyl fermions were observed in transport experiments in the fully-polarized magnetic phase of CeSb~\cite{GuoNPJQM2017}.
However, the Weyl fermion bands have not yet been directly resolved in any phase of Ce$X$, because the TRS-broken phases exist only under external magnetic field, which precludes the use of angle-resolved photoemission spectroscopy (ARPES).
In cerium monopnictides, it remains crucial to measure the surface magnetic order, its associated band splitting, and its orbital contributions, which could influence the Fermi arcs and their connectivity to Weyl cones~\cite{MoraliScience2019}.  
Furthermore, characterizing the interplay between magnetic order and Kondo physics is essential to understand the possible emergence of heavy fermions and flat bands~\cite{JangSciAdv2019}.

\begin{figure}[ht]
    \includegraphics[width=\columnwidth]{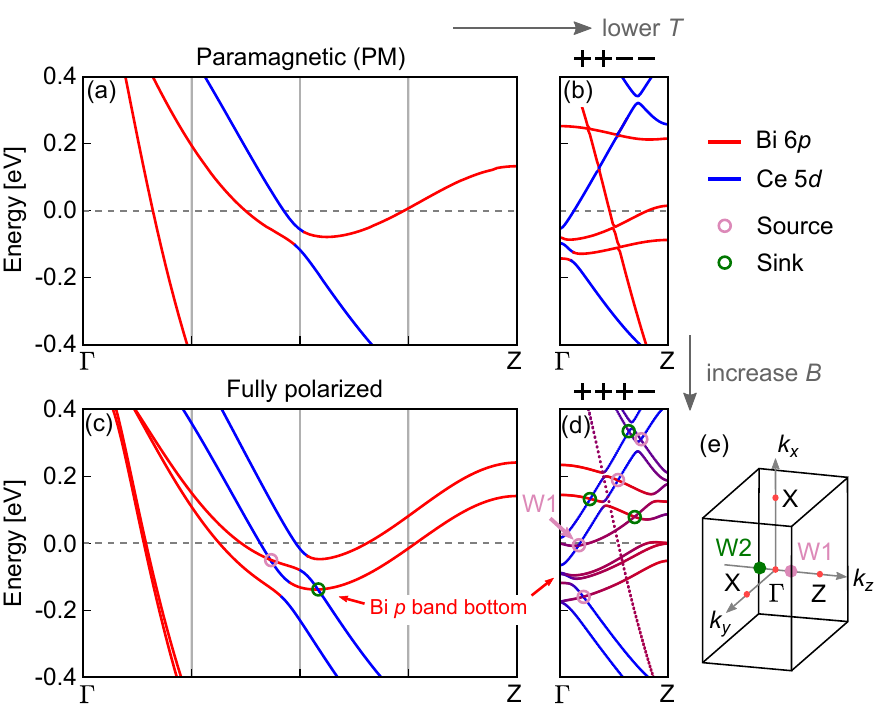}  
    \caption{DFT band structure along the $\Gamma-Z$ direction in the (a) paramagnetic; (b) antiferromagnetic \ppmm; (c) fully-polarized \pppp; and (d) ferrimagnetic \pppm\ phases. Light purple and dark green circles indicate locations of Weyl nodes. W1 and W2 indicate the Weyl nodes closest to the measured Fermi energy ($\EF$) in the ferrimagnetic phase. The calculated bands in each panel have been rigidly shifted by (a) $+110$ meV, (b) $-120$ meV, (c) $-110$ meV, and (d) $-55$ meV to better match our QPI experiment. (e) Full 3-dimensional Brillouin zone (BZ) in the \pppm\ phase.} 
    \label{fig:DFT}
\end{figure}

\ptitle{Here we show...}
Here we use spin-polarized (SP) scanning tunneling microscopy (STM) and spectroscopy (STS) to image the energy-resolved surface magnetic order on CeBi, at low temperature with applied magnetic field $B$. We extend previous density functional theory (DFT) calculations of Weyl nodes in the high-$B$ fully-polarized phase~\cite{HuangPRB2020}, to predict additional Weyl nodes near the Fermi level $\EF$ in the intermediate-$B$ ferrimagnetic phase. We therefore focus our experiments at $B=3$ T, where we image the expected \pppm\ pattern of the spin orientation of the $f$ orbitals on the Ce sites, but only above $\EF$. Surprisingly, we observe the same magnetic pattern on the Bi sites below $\EF$. The induced magnetic moments on the Bi $p$ states, co-aligned with the adjacent Ce $f$ states, indicate $p - f$ orbital hybridization, commonly referred to as $p-f$ mixing~\cite{TakahashiJPCSSP1985,LiPRB2019}. For energies closer to $\EF$ we observe suppression of the \pppm\ spin polarization, coinciding with a Fano resonance in our measured conductance ($dI/dV$), further supporting $p-f$ hybridization. These observations suggest a competition between the mechanism inducing the co-aligned (i.e.\ ferromagnetically aligned) moments on the Bi $p$ states and the antiferromagnetic Kondo screening. Finally, we present quasiparticle interference (QPI) measurements showing a $\sim$100 meV splitting of the Bi $p$-band, validating our DFT calculations that show the crossing of the Bi $p$ and Ce $d$ bands to form Weyl nodes close to $\EF$. Our QPI suggests a flattening of the mixed-character $p-d$ band that forms the Weyl cones.

\section{Methods}
 
\ptitle{DFT} We calculated the bulk band structure of CeBi using the generalized-gradient approximation (GGA) as implemented in the all-electron code WIEN2K~\cite{wien2k,*wien2k2}, with the augmented-plane-wave + local-orbitals (APW+lo) basis set. In the paramagnetic phase we treated the Ce 4$f$ orbitals as core electrons while in the magnetic phases we incorporated the Hubbard Coulomb interaction on the Ce $4f$ electrons, with $U=7.9$ eV and $J=0.69$ eV chosen to make the Ce $4f$ energy level qualitatively consistent with ARPES measurements on CeBi~\cite{LiPRB2019}. We included spin-orbit coupling in all calculations.

\ptitle{Experiments} Single crystals of CeBi were grown by the  self-flux method~\cite{Kuthanazhi2021}. We cooled the crystals in zero field, cleaved them in cryogenic ultra-high vacuum at $\sim30$ K to expose a neutral (010) surface before imaging them at $T=4.6$ K. We prepared non-magnetic PtIr STM tips by \textit{ex situ} mechanical sharpening, then \textit{in situ} field emission on Au foil. We obtained spin-polarized tips by gently dunking them into  the sample to pick up a few atoms of magnetic material~\cite{KroenleinPRL2018,EnayatScience2014,LothNatPhys2010}.

\section{Density functional theory calculations} 

\ptitle{DFT predictions} Figure \ref{fig:DFT} shows the folding and splitting of the two Bi $p$ bands and the Ce $d$ band that cross $\EF$, as $T$ is lowered and $B$ is increased~\cite{repo}. Our calculations predict an induced magnetic moment of $\sim 0.01 \mu_\mathrm{B}$ on the Bi $p$ states in both the ferrimagnetic and the fully-polarized phase.
We also computed the Berry curvature at each band crossing in the fully-polarized and ferrimagnetic phases, and circled the sinks and sources that constitute Weyl nodes (see also Fig.\ \ref{fig:Berry})~\cite{HuangPRB2020}. In the fully-polarized phase we find a band splitting of $\sim$100 meV that generates Weyl nodes near the calculated $\EF$, however we caution that the true $\EF$ may be significantly shifted in low-carrier-density Kondo materials. The ferrimagnetic phase shows comparable band splitting, but is advantageous because its band folding \textit{generates additional Weyl nodes}. The predicted ferrimagnetic-phase Weyl nodes are well-spaced over a larger energy range, making the Weyl phase and its low energy signatures more robust to band bending. This circumvents the problem of ill-defined chirality that arises from scattering between multiple degenerate Fermi arcs~\cite{XuScience2015} or from Fermi surfaces encompassing multiple Weyl points~\cite{ArnoldNatCom2016}. We therefore focus our attention experimentally on the ferrimagnetic phase for the remainder of this work. 

\begin{figure}[t]
    \includegraphics[width=\columnwidth]{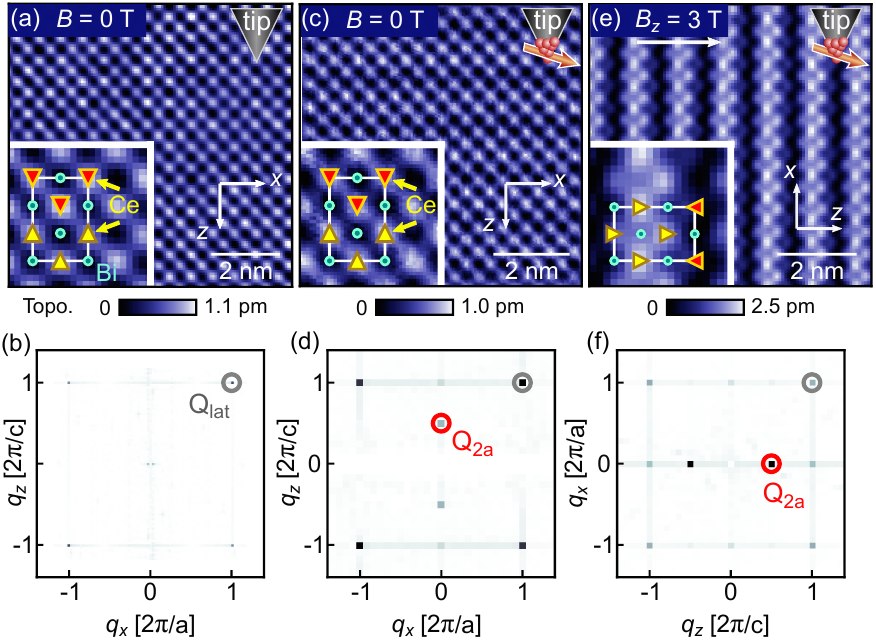}
    \caption{(a) Topography of CeBi measured with spin-degenerate PtIr tip at zero field. (Sample bias $\Vs = 400$ mV, current setpoint $\Is = 200$ pA.)  (b) Fourier transform (FT) of (a) shows Ce sublattice peaks at $\Qlat$, but no magnetic peaks. (c,d) Topography and FT with spin-polarized tip (obtained by dunking to pick up magnetic material) show new structure and dominant Bragg peaks at $\Qta$ (red circle), consistent with the expected bulk antiferromagnetic \ppmm\ phase that breaks the cubic symmetry even in zero applied field. ($\Vs = 400$ mV, $\Is = 150$ pA.) (e,f) By applying a horizontal field of $B_z=3$ T orthogonal to the \ppmm\ wavevector of (c,d), the spin-polarized topography and FT show reoriented structure and Bragg peaks at $\Qta$ (red circle) consistent with the expected bulk ferrimagnetic \pppm\ phase  ($\Vs = 100$ mV, $\Is = 500$ pA, recorded in vicinity of (c) and with the same spin-polarized tip). Insets in panels (a),(c) and (e) show overlaid Bi (cyan) and Ce (yellow and red) atoms, with circles denoting expected non-magnetic orbitals and triangles denoting the expected spin orientations of the Ce $f$ orbital, from neutron scattering \cite{BartholinJPC1979}.}
     \label{fig:topos}
\end{figure}

\begin{figure}[ht]
    \includegraphics[width=\columnwidth]{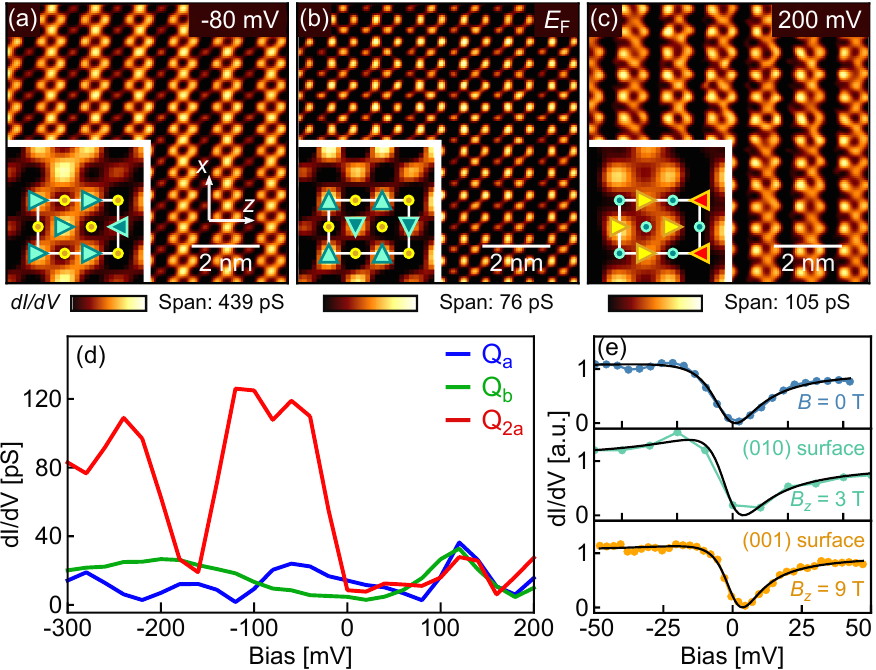}
    \caption{Spin-polarized conductance maps at $B_z=3$ T of the (a) filled, (b) Fermi level, and (c) empty states. ($\Vs = 300$ mV, $\Is = 4$ nA, bias modulation $\Vrms = 7.1$ mV.) The energies of the three maps highlight the filled and Fermi level Bi $p$ orbitals, and the unoccupied Ce $d$ orbitals, respectively. As in Fig.\ \ref{fig:topos}(a,c,e), insets show overlaid Bi (cyan) and Ce (yellow and red) atoms, but here the triangles denote the observed \textit{induced} spin orientations on the Bi $p$ (a,b) and Ce $d$ (c) orbitals. 
    (d) Energy-dependent intensity of magnetic Bragg peaks (as defined in Fig.\ \ref{fig:phase_diagram}) on the (010) surface of the \pppm\ phase of CeBi. The dip in $\Qta$ around $-150$ mV corresponds to the bottom of the outer Bi $p$ band, which is marked by a red arrow in Fig.\ \ref{fig:DFT}(d).  (e) Background-subtracted, spatially-averaged $dI/dV$ curves, fit to Fano line shape (see Figs.\ \ref{fig:kondo} and \ref{fig:kondo_linecut}).}
     \label{fig:dIdV}
\end{figure}

\begin{figure}[ht!]
    \includegraphics[width=\columnwidth]{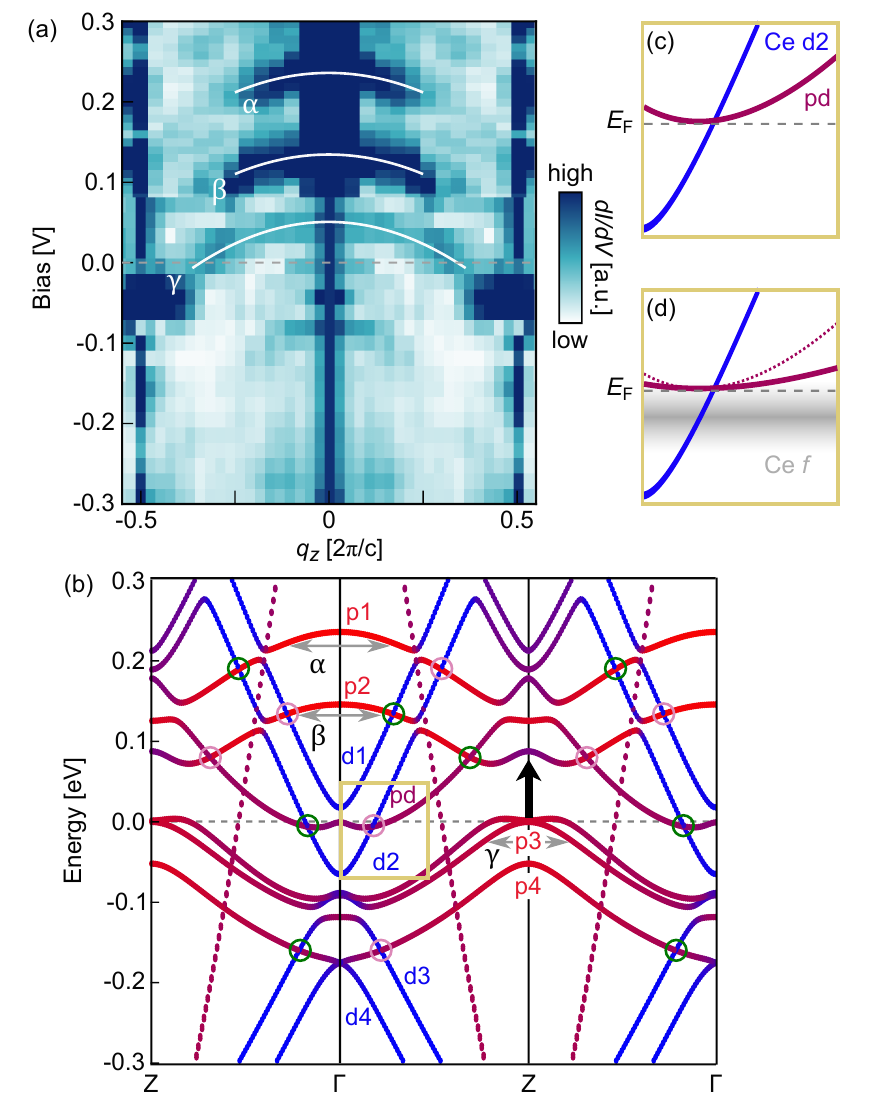}
    \caption{(a) Quasiparticle interference (QPI) intensity along the $\Gamma-Z$ direction of the \pppm\ phase at $B_z=3$ T (see Fig.\ \ref{fig:QPI_app} for details). (b) DFT band structure from Fig.\ \ref{fig:DFT}(d) in an extended zone scheme. Predicted Weyl nodes are circled in light purple and dark green.
    The dominant features in the QPI data, labeled by three white arcs $\alpha$, $\beta$, and $\gamma$ in (a), match the three intraband scattering processes $p1$, $p2$, and $p3$ marked as gray arrows on the DFT in (b). The predominance of these three $p$-orbital band segments in the QPI signal is consistent with a prior observation that $p$ orbitals extend farther from the surface than $d$ orbitals, and are more accessible to the STM tip~\cite{ZhengNanoLett2016}. Our QPI data also shows that the $p3$ segment of this folded outer $p$ band is shifted up by $\sim70$ meV (black arrow) with respect to the $p1$ and $p2$ segments, consistent with band flattening due to strong electron correlations.
    (c) DFT band structure close to tilted (type II) Weyl cone at $\EF$. (d) Schematic showing the renormalization (flattening) of the Bi $6p$ band upon hybridization with the Ce $4f$ states~\cite{KasuyaPhysBCondMatt1993}.
    Dashed red line denotes the non-renormalized $p$ band. Gray shading denotes the renormalized $f$ state forming our observed Kondo resonance, in agreement with ARPES~\cite{JangSciAdv2019}. Dashed black line indicates experimental $\EF$.}
     \label{fig:QPI}
\end{figure}

\section{Experimental Results} 

\ptitle{Topography shows spin-polarized STM} Figure\ \ref{fig:topos}(a) shows a topography acquired with a non-magnetic PtIr tip. The corresponding Fourier transform (FT) in Fig.\ \ref{fig:topos}(b) shows four peaks at $\Qlat = (\pm 1,\pm 1)$, arising from the Ce sublattice (see Fig.\ \ref{fig:LaDoping} ). After gently dunking the tip into the sample, the new topography in Fig.\ \ref{fig:topos}(c) shows additional structure, suggesting that the tip has picked up magnetic material, leading to a spin-polarized tunneling current~\cite{WiesendangerRMP2009, EnayatScience2014}. The magnetic structure manifests in the FT in Fig.\ \ref{fig:topos}(d) as two dominant new peaks at $\Qta = (0,\pm\frac{1}{2})$, consistent with the expected bulk antiferromagnetic \ppmm\ phase that breaks the cubic symmetry of CeBi even in the absence of applied $B$ [see also Fig.~\ref{fig:phase_diagram}(c)].  To verify the SP nature of the tip,
we applied an in-plane field $B_z=3$ T, perpendicular to the zero-field \ppmm\ order, to rotate the magnetic ordering vector of the sample by ninety degrees into the expected \pppm\ phase, as shown in Fig.\ \ref{fig:topos}(e)-\ref{fig:topos}(f) [see also Fig.\ \ref{fig:phase_diagram}(d)]. 
Although the tip spin may also realign due to applied $B$, our observations in Fig.\ \ref{fig:topos}(c) and \ref{fig:topos}(e) show that the tip has a component that is co-aligned with the sample magnetization in both cases. This confirms two essential requirements for our study: Our tip is sensitive to the expected magnetic order of CeBi, and we can tune that magnetic order by applying magnetic field. 

\ptitle{$dI/dV$ suggests $p-f$ hybridization} To determine the energy-resolved orbital character of the spins in the ferrimagnetic \pppm\ phase, we mapped the spin-polarized differential conductance, $dI/dV$. Figure\ \ref{fig:dIdV}(a)-\ref{fig:dIdV}(c) shows SP-$dI/dV$ maps from energies below, at, and above $\EF$. Away from $\EF$, we find a high SP-conductance for three neighboring vertical columns and a low SP-conductance for one column, as expected in the \pppm\ ferrimagnetic phase. Surprisingly, comparison of the simultaneously-acquired images in Figs.\ \ref{fig:dIdV}(a) and \ref{fig:dIdV}(c) reveals that the magnetic contrast has shifted from the Ce lattice sites above $\EF$ to the Bi lattice sites below $\EF$. These induced magnetic moments on the Bi $p$ sites are co-aligned with the underlying magnetic pattern of the localized, unrenormalized Ce $f$ states, which have been observed $\sim 3 $ eV below $\EF$ by ARPES~\cite{LiPRB2019}. Our evidence of $p - f$ orbital hybridization agrees with our DFT prediction of a Bi $p$ magnetic moment of $\sim 0.01 \mu_\mathrm{B}$.

\ptitle{$\EF$ suppression}
We plot the intensity of the magnetic Bragg peaks vs.\ energy in Fig.\ \ref{fig:dIdV}(d). The $\Qta$ peak is dominant at negative energies, but strongly suppressed at $\EF$, and recovers only weakly above $\EF$. This evolution is also apparent in the colorscale spans of the SP-$dI/dV$ maps. While the \pppm\ pattern of induced magnetic moments is dominant at $\Vs = -80$ and $+200$ meV in Figs.\ \ref{fig:dIdV}(a) and \ref{fig:dIdV}(c), the map at $\EF$ in Fig.\ \ref{fig:dIdV}(b) shows a different motif with maxima on every second Bi atom.  We speculate that this \pmpm\ structure may be caused by a residual out-of-plane ordering of surface spins (see Fig.\ \ref{fig:canting}). The suppression of the $\Qta$ intensity also coincides with a $180^\circ$ phase flip of the real-space pattern (see Fig.\ \ref{fig:LaDoping}).

\ptitle{Kondo resonance} 
Kondo screening can arise from $p - f$ hybridization, as suggested by previous Hall effect and ARPES measurements~\cite{KitazawaJOMMM1985,LiPRB2019}, so we search for its possible signatures in $dI/dV$ spectroscopy.
Fig.\ \ref{fig:dIdV}(e) shows three spatially-averaged $dI/dV$ spectra around $\EF$ in the \ppmm, \pppm, and fully-polarized phases. All spectra have a similar shape with a shoulder around $-18$ meV and a dip near $\EF$, characteristic of the asymmetric Fano line shape describing a Kondo resonance, 
\begin{equation}
    F(E) \propto \frac{[q+(E-E_0)/\Gamma]^2}{1+[(E-E_0)/\Gamma]^2}\,,
\label{eq:Fano}
\end{equation}
where $E_0$ is the energy of the localized many-body resonance (which we find to be consistent with $\EF$; see Table \ref{tab:surf}), $q$ is the Fano factor that describes the tunnelling ratio between the localized $f$ state and the itinerant conduction electrons, and $\Gamma$ is proportional to the hybridization between the localized and itinerant states~\cite{FanoPR1961}. 
Similar line shapes have been observed in other Kondo lattice systems such as YbRh$_2$Si$_2$~\cite{ErnstNature2011}, URu$_2$Si$_2$~\cite{SchmidtNature2010}, and SmB$_6$~\cite{PirieNatPhys2020}. In all three magnetic phases of CeBi, we find $\Gamma \sim 10$ meV, consistent with a resistivity upturn at $\sim$100 K, above the N\'eel temperature~\cite{Kuthanazhi2021}. However, the large residual $dI/dV$ at the Fano minimum (see Fig.\ \ref{fig:kondo}) suggests that only $5-10\%$ of the conduction electrons participate in Kondo screening in CeBi, consistent with the local-moment-like behavior of the Ce sublattice~\cite{BartholinJPC1979, Kuthanazhi2021}.

\ptitle{Kondo screening is from $p$ not $d$} 
In CeBi the itinerant electrons closest to $\EF$ are of both Ce $5d$ and Bi $6p$ character, so we do not know \textit{a priori} which itinerant states participate in the partial Kondo screening of Ce $f$ moments. 
However, the disappearance of \pppm\ order from the Bi sites at $\EF$ is consistent with the involvement of Bi $p$ states in Kondo singlet formation. Furthermore, the Kondo resonance is facilitated by the shared symmetry of the Bi $6p$ orbitals and the Ce $f_{5/2}\Gamma_8$ multiplet~\cite{HeerJPCSSP1979,KasuyaPhysBCondMatt1993}, while $d-f$ hybridization is forbidden on-site. We thus conclude that the Bi $6p$ states are the primary conduction electrons that couple to the Ce $4f$ moments and participate in the formation of Kondo singlets.

\ptitle{Competing interactions}
Fig.\ \ref{fig:dIdV} highlights several competing interactions in CeBi. First, the Fano lineshape in our $dI/dV$ spectra suggests a Kondo resonance in which some conduction electrons \textit{anti-align} with and partially screen the local moments. However, the observed shift of the same \pppm\ order from the Ce sites in Fig.\ \ref{fig:dIdV}(c) to the Bi sites in Fig.\ \ref{fig:dIdV}(a) demonstrates that the induced magnetic moments on the Bi $p$ states are \textit{co-aligned} with the local $f$ moments. These opposite magnetic interactions compete for the same $p$ states. Second, long-range order can arise when localized moments couple to each other via the polarized conduction electrons, through the Ruderman-Kittel-Kasuya-Yosida (RKKY) interaction. The competition between the screening of the local moments (Kondo) and formation of long-range magnetic order (RKKY) is typically determined by both the conduction ($c$) electron density and the coupling $J_{f-c}$~\cite{DoniachPhysicaB1977,Coleman2015}. However, we observe long-range order in Figs.\ \ref{fig:dIdV}(a) and \ref{fig:dIdV}(c) coexisting with the Kondo resonance in Fig.\ \ref{fig:dIdV}(e).  Our data suggests that these competing interactions in CeBi each dominate at separate \textit{energies}, in contrast with CeSb where the simultaneous observation of Kondo screening and long-range order has been explained by phase separation in \textit{momentum} space~\cite{JangSciAdv2019}.

\ptitle{DFT reproduces band splitting} To determine the effect of $p-f$ hybridization on the predicted Weyl fermion bands in the CeBi \pppm\ phase, Fig.\ \ref{fig:QPI}(a) shows our quasiparticle interference (QPI) measurement of the band dispersion along $\Gamma-Z$. The dominant QPI features show excellent agreement with the calculated bands of majority Bi $p$ orbital character shown in Fig.\ \ref{fig:QPI}(b). From this band assignment, we make several observations. First, the QPI-observed $\alpha$ and $\beta$ bands support the DFT-predicted $\sim$100 meV splitting of the hole-like Bi $p$ band around $\Gamma$ into $p1$ ($\alpha$) at 230 meV and $p2$ ($\beta$) at 120 meV, consistent with the formation of induced magnetic moments on the Bi $p$ sites by orbital hybridization. 
Second, the QPI-observed $\gamma$ band can be attributed to scattering across the BZ boundary between the folded $p3$ portion of the same Bi $p$ band. However, the QPI-observed $\gamma$ band is $\sim70$ meV higher than the corresponding DFT-predicted $p3$ band, supporting a scenario of $p-f$ hybridization that renormalizes (flattens) the $p$ band and the band of mixed $pd$ character that makes up half of the Fermi-level Weyl cone. We thus infer that the Weyl cones of CeBi are strongly renormalized compared to the DFT calculations, consistent with the enhanced effective mass of $4.3 m_e$ observed in quantum oscillation experiments on CeSb~\cite{SettaiJPSJ1994}. DFT calculations notoriously underestimate band-renormalization effects in strongly interacting materials, so our QPI experiment serves as a crucial reality check.\\

\section{Conclusion}
\ptitle{Implications for Weyl}
Weyl cones are expected to be robust under renormalization, which can spread their position in momentum space but not lift the degeneracy of the requisite crossings~\cite{WitczakARCMP2014}. Neither $p-f$ nor $d-f$ hybridization yields an exact realization of the proposed Weyl-Kondo semimetal~\cite{LaiPNAS2018}, in which the $f$ electrons are directly involved in the formation of the Weyl cones. However, our work shows (\textit{i}) induced magnetic moments on the Bi $p$ states, co-aligned with the local Ce $f$ moments, from spin-polarized $dI/dV$ images of the \pppm\ order on the Bi sites;
(\textit{ii}) $p-f$ hybridization from $dI/dV$ spectroscopy of the Fano resonance; (\textit{iii}) $\sim$100 meV band splitting from QPI that confirms $p-f$ hybridization and TRS breaking; (\textit{iv}) and band flattening from QPI. This comprehensive evidence supports a consistent picture of CeBi as a strongly interacting magnetic Weyl semimetal.

\section*{Acknowledgements}

We thank Jason Hoffman, Robert Jan-Slager, Daniel Mazzone for insightful discussions. Experimental and theoretical work was supported by the Center for the Advancement of Topological Semimetals (CATS), an Energy Frontier Research Center funded by the U.S. Department of Energy (DOE), Office of Science, Basic Energy Sciences (BES) through the Ames Laboratory under its Contract No. DE-AC02-07CH11358. H.P. was funded by the Gordon and Betty Moore Foundation's EPiQS Initiative through Grant GBMF4536. C.E.M. was supported by the Swiss National Science Foundation under fellowships P2EZP2\_175155 and P400P2\_183890. The theory work was carried out under the auspices of the U.S. DOE National Nuclear Security Administration under Contract No. 89233218CNA000001. C.L. was supported by Los Alamos National Laboratory (LANL) LDRD Program. The theory was also supported in part by the Center for Integrated Nanotechnologies, a DOE BES user facility, in partnership with the LANL Institutional Computing Program for computational resources.\\

All data underlying Figs. \ref{fig:phase_diagram}-\ref{fig:interpol} can be accessed in Ref.~\cite{repo}.

\vspace{-2mm}
\appendix
\section{Surface atom and spin identification}

\begin{figure*}[t!]
    \includegraphics[width=\textwidth]{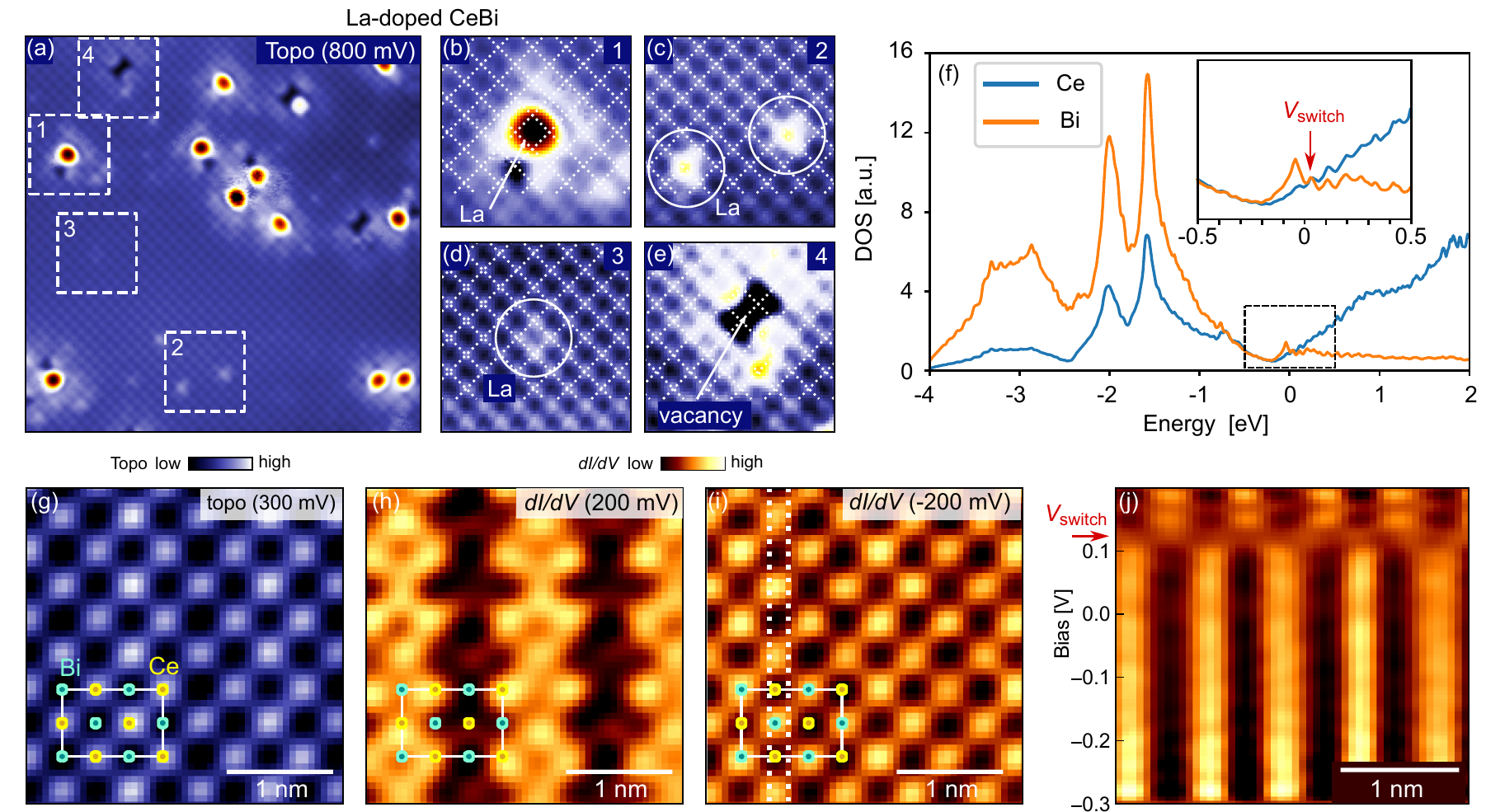}
    \caption{
    (a-e) Topography of CeBi with 2.5\% (nominal) La dopants expected at the Ce sites, recorded with sample bias $\Vs = 800$ mV and tunneling current setpoint $\Is = 600$ pA. The spacing of the visible atomic lattice (periodic bright spots) in a clean region is 4.6 \AA, consistent with x-ray diffraction measurements of the lattice constant \cite{HulligerJLTP1975}. (b) Zoom around a typical La adatom that is centered between bright lattice spots. (c) Two La dopants in the top surface layer, located on the bright lattice sites. (d) La in the subsurface layer, laterally centered on a dark lattice site of the top surface layer. (e) Vacancy of two bright lattice sites of the top surface layer. All four observations in (b-e) suggest that at the large positive sample bias of $\Vs = 800$ mV, we observe the Ce sublattice.  (f) DFT-calculated density of states (DOS) of bulk CeBi in the ferrimagnetic \pppm\ state, resolved according to elemental contribution. At negative energies, the DOS of Bi dominates while on the positive side, Ce dominates. (g-j) Topography and $dI/dV$ maps of nominally pristine (undoped) CeBi, to identify the shift of dominant sublattice depending on sample bias voltage. (g) Topography recorded at $\Vs = 300$ mV. (h-j) Simultaneously recorded $dI/dV$ maps at $+200$ mV and $-200$ mV, respectively. (j) $dI/dV$ linecut spatially averaged between the vertical dotted lines in (i), illustrating the spatial switching of highest-conductance location at $V_\mathrm{switch} \sim 0.1$ V. Average magnitude of horizontal rows in (j) has been normalized for visual purposes. All maps in (g-j) have been simultaneously recorded at $B_z = 3$ T with $\Vs = 300$ mV, $\Is = 4$ nA, and $\Vrms = 7.1$ mV. 
    }
    \label{fig:LaDoping}
\end{figure*}

\begin{figure*}[h]
    \includegraphics[width=0.9\textwidth]{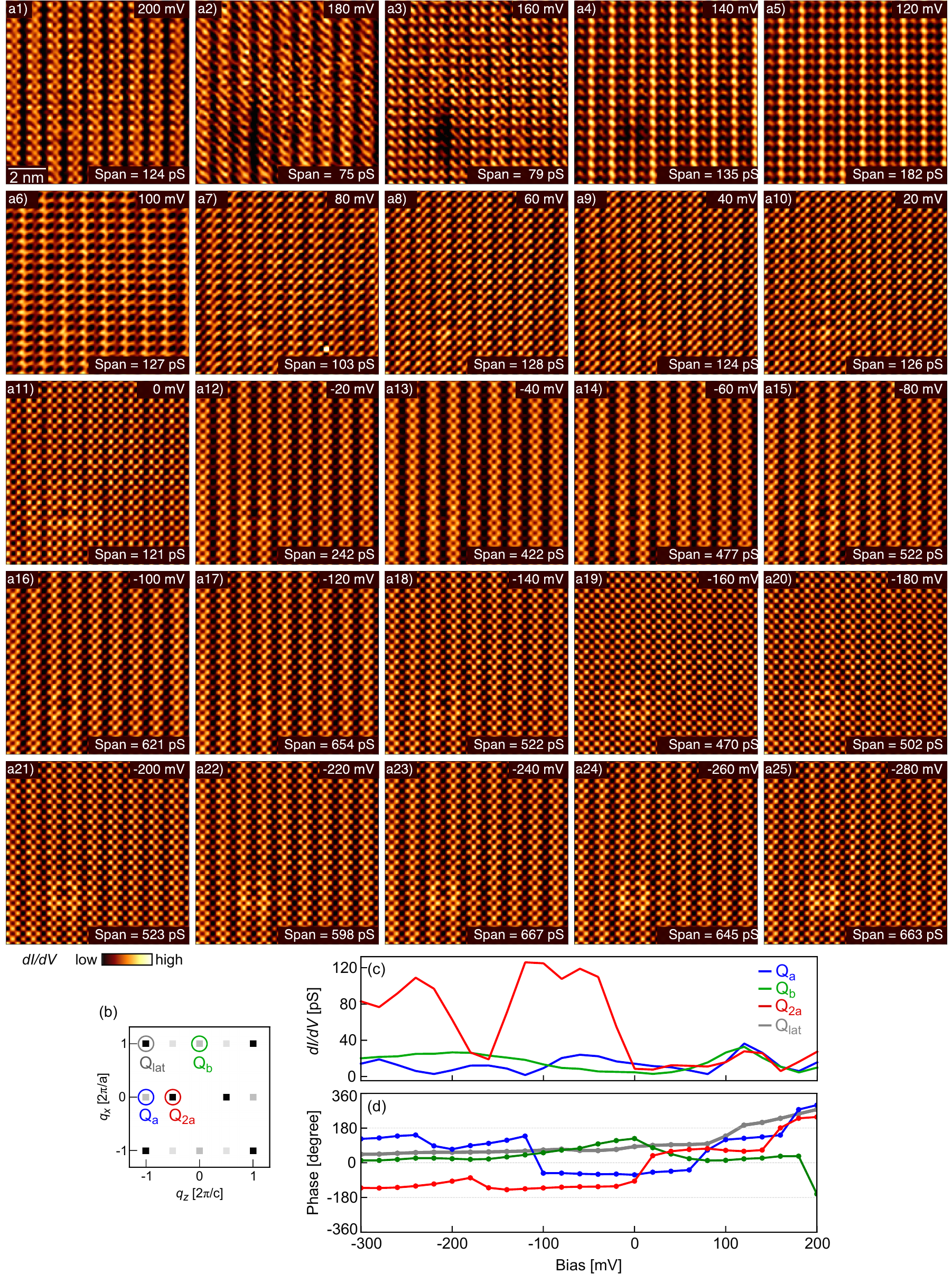}
    \caption{(a1-a25) All energy layers of the conductance map presented in Fig.\ \ref{fig:dIdV}, showing the (010) surface of the \pppm\ phase at $B_z=3$ T. Sample bias is indicated in top right corner of each layer. Setup parameters are $\Vs = 300$ mV, $\Is = 4$ nA, $\Vrms = 7.1$ mV. (b) Simulated Fourier transform of \pppm\ magnetic order on the (010) surface with magnetic ($\Qa$,$\Qb$,$\Qta$) and structural lattice ($\Qlat$) Bragg peaks as indicated. (c-d) Energy-dependent (c) intensity and (d) phase of magnetic and structural Bragg peaks, calculated by Fourier transforming each measured conductance map in panels (a1-a25).}
    \label{fig:allenergies}
\end{figure*}

\begin{figure}[h]
    \includegraphics[width=\columnwidth]{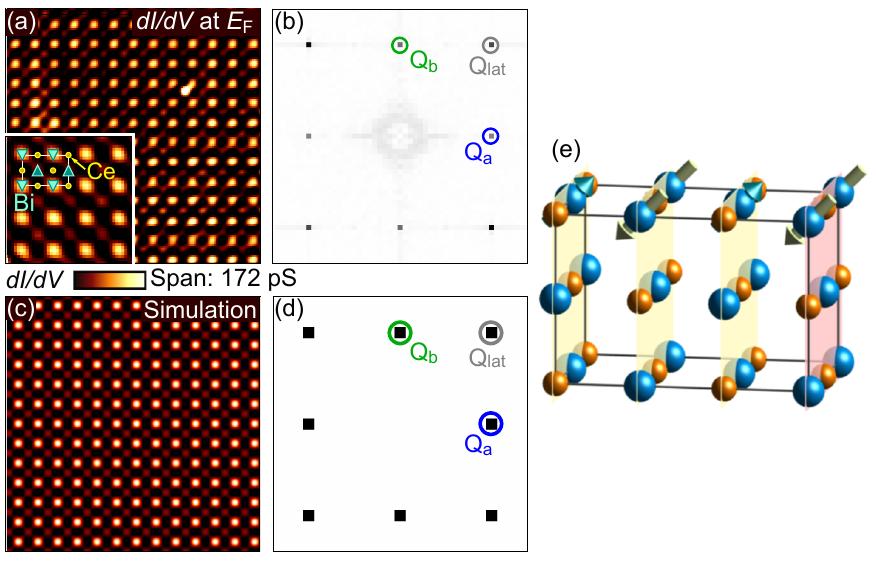}
    \caption{
    (a) Spin-polarized $dI/dV$ map at $B_z=3$ T, at the Fermi level, shows alternating magnetic \pmpm\ contrast, which is different from the expected in-plane \pppm\ magnetic order from the phase diagram of Fig.\ \ref{fig:phase_diagram}(d). 
    This map was extracted from the same dataset presented in Fig.\ \ref{fig:LaDoping}(g-j), with $\Vs = 300$ mV, $\Is = 4$ nA, $\Vrms = 7.1$ mV. Although this map and Fig.\ \ref{fig:allenergies}(a11) were recorded with identical parameters, the tip termination had a different direction of the magnetic moment (spin-DOS), so the real space images do not appear identical, but $\Qa$ and $\Qb$ are prominent in the Fourier transform in both cases. (b) Corresponding Fourier transform shows that the intensity of the $\Qta$ magnetic Bragg peaks is suppressed, indicating suppression of the bulk \pppm\ magnetic order on the Bi $p$ states due to the (partial) formation of Kondo singlet states.  (c-d) Simulation of the \pmpm\ magnetic order. (e) Possible scenario to explain the \pmpm\ order at $\EF$: spins on top (010) surface may have a residual out-of-plane component, which would produce the intensity pattern in our simulation (c-d).}
    \label{fig:canting}
\end{figure}

To identify the atomic sublattice imaged in Figs.\ \ref{fig:topos} and \ref{fig:dIdV}, we investigated La-doped CeBi samples, where La is expected to replace Ce. Measurements and DFT calculations in Fig.\ \ref{fig:LaDoping} show that at large positive bias voltage we are tunneling predominantly into Ce sites, while below $\sim 0.1$ V we are tunneling predominantly into Bi sites. Fig.\ \ref{fig:allenergies} shows the full energy dependence of the spin-polarized $dI/dV$ maps in Fig.\ \ref{fig:dIdV}. Fig.\ \ref{fig:canting}(a) shows a second spin-polarized $dI/dV$ map at the Fermi level, acquired with identical measurement parameters as Fig.\ \ref{fig:allenergies}(a11) but with different tip termination. Together, Figs.\ \ref{fig:allenergies}(a11) and \ref{fig:canting}(a) suggest a canting of the surface spins.

The Fermi level $dI/dV$ map in Fig.\ \ref{fig:canting}(a) shows a pattern with maxima on every second Bi atom, which is different from the \pppm\ pattern of the bulk magnetic moments expected from the phase diagram of Fig.\ \ref{fig:phase_diagram}(d). This Fermi level $dI/dV$ corresponds to dominant $\Qa$ and $\Qb$ magnetic Bragg peaks, as shown in Fig.\ \ref{fig:canting}(b). Figure \ref{fig:allenergies}(c) shows a corresponding suppression of the $\Qta$ Bragg peak associated with the \pppm\ order for energies close to the Fermi level. The suppression of the $\Qta$ intensity also coincides with a $180^\circ$  phase flip of the real-space pattern in Fig.\ \ref{fig:allenergies}(a). The suppression of $\Qta$ at the Fermi level implies that the \pmpm\ pattern seen at that energy is not connected to the \pppm\ order, but of different origin. We suggest that this \pmpm\ structure might be caused by a residual out-of-plane ordering of surface spins, as sketched in Fig.\ \ref{fig:canting}(e), and simulated in Figs.\ \ref{fig:canting}(c)-(d).

\section{Kondo lineshape}

\begin{figure}[h!]
    \includegraphics[width=1\columnwidth]{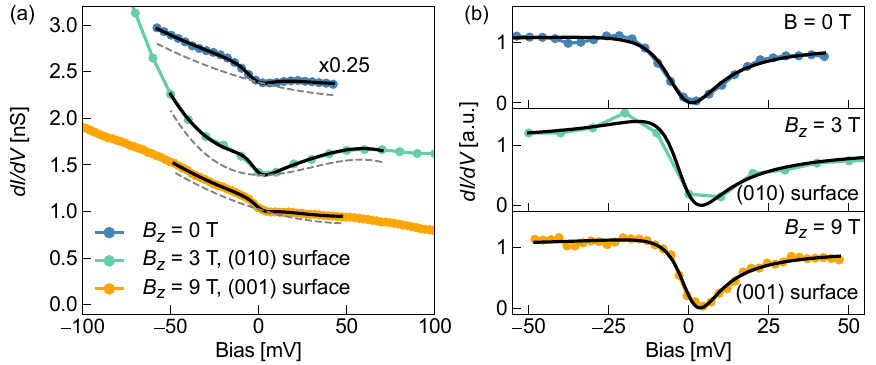}
    \caption{\textbf{Kondo resonance in CeBi}
    (a) Spatially-averaged $dI/dV$ spectra around $\EF$ with varying applied $B$. Acquisition parameters are: (blue, $B=0$~T) $\Vs = 50$~mV, $\Is = 1$~nA, $\Vrms = 2.82$~mV; (green, $B_z=3$~T) $\Vs = 300$ mV, $\Is = 2$~nA, $\Vrms = 7.1$~mV; (orange, $B_z = 9$~T), $\Vs = 200$~mV, $\Is = 0.4$~nA, $\Vrms = 1.8$~mV. Blue and green curves are spatially averaged spectra from DOS maps recorded on the (010) surface with a spin-polarized tip, while orange curve is nominally a point spectrum with a non-spin-polarized tip on the (001) surface, it was recorded over an extended time, so it is effectively spatially averaging due to lateral piezo drift in nm range. Black lines show the fit of a Fano lineshape (Eqn.\ \ref{eq:Fano}) on top of a polynomial background (parabolic for 0~T and 9~T, but third order polynomial for 3~T data), depicted by gray dashed lines.
    (b) Background-subtracted $dI/dV$, overlaid with fits to Fano lineshape, as shown in Fig.\ \ref{fig:dIdV}(e).}
    \label{fig:kondo}
\end{figure}

\begin{figure}[h!]
    \includegraphics[width=1\columnwidth]{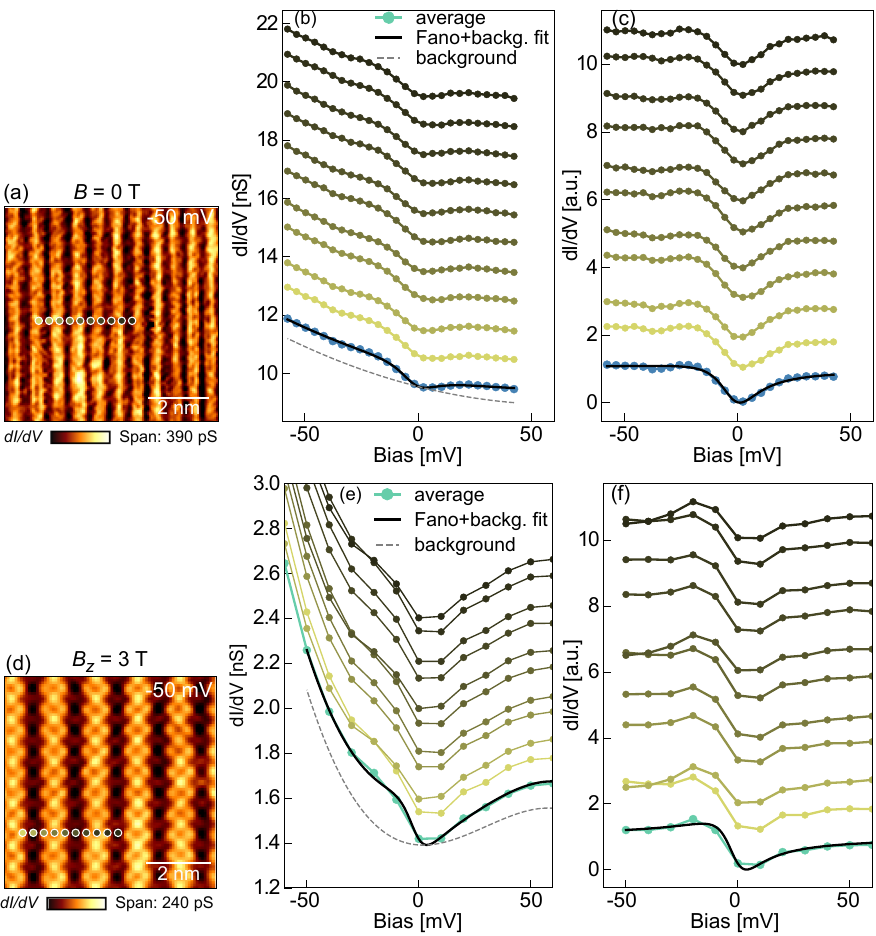}
    \caption{\textbf{Point conductance spectra at 0 T and 3 T in CeBi}
    (a) Conductance map recorded at -50 mV bias ($B=0$ T, $\Vs = 50$ mV, $\Is = 1$ nA, $\Vrms = 2.82$ mV) indicating the location of the point spectra presented in (b). (b) Raw point spectra vertically offset for clarity. Blue spectrum is spatial average within the field of view in panel (a), gray dashed line is polynomial background fit, and black line is Fano lineshape fit plus background. (c) Point spectra vertically offset for clarity after background subtraction, indicating Fano-like lineshape at all locations. (d)-(f) Similar image and point spectra recorded at $B_z = 3$ T ($\Vs = 300$ mV, $\Is = 2$ nA, $\Vrms = 7.1$ mV). }
    \label{fig:kondo_linecut}
\end{figure}

Figure \ref{fig:kondo}(a) shows the spatially-averaged raw $dI/dV$ spectra around $\EF$ in the \ppmm, \pppm, and fully-polarized phase from which we subtracted the background (gray dashed lines) to obtain the Fano lineshapes shown in Fig.\ \ref{fig:dIdV}(e), and re-displayed here in Fig.\ \ref{fig:kondo}(b).
The qualitative features of the Fano lineshape, with a shoulder around $-18$ meV and a dip near $\EF$, are apparent in the raw spectra. But to quantify the Fano lineshape parameters in Table \ref{tab:surf}, we fit Eqn.\ \ref{eq:Fano} added to a polynomial background.  By comparing the residual conductance at $\EF$ with the shoulder around $-20$ meV in Fig.\ \ref{fig:kondo}(a), it can be seen that only $5-10\%$ of the conduction electrons participate in the Kondo screening. 
Furthermore, the point spectra in Fig.\ \ref{fig:kondo_linecut} recorded in the antiferromagnetic ground state ($B = 0$ T) and in the ferrimagnetic state ($B_z = 3$ T) all show a Fano-like lineshape with only slight spatial variation at larger binding energies due to the magnetic order. 

\begin{table}[h]
\caption{\label{tab:surf} Fano line shape parameters determined by a least-square fit. The Kondo temperature is estimated as $T_K=\Gamma/k_B$ where $k_B$ is the Boltzman constant. }
\begin{tabular}[c]{c | r c l | r c l | c | c}
   \hline
   & \multicolumn{3}{c|}{$\Gamma$ [meV]} & \multicolumn{3}{c|}{$E_0$ [meV]} & $q$ & $T_K$ [K]
  \\ \hline
   $B = 0$ T & 10.2 & $\pm$ & 4 & $-1.9$ & $\pm$ & 4 & $-0.3$ & 118.37
  \\ \hline
   $B_z = 3$ T & 8.8 & $\pm$ & 10 & $-1.6$ & $\pm$ & 10 & $-0.6$ & 102.12
  \\ \hline
   $B_z = 9$ T & 7.7 & $\pm$ & 2.5 & 0.8 & $\pm$ & 2.5 & $-0.4$ & 98.35
  \\ \hline
\end{tabular}
\end{table}

\clearpage

\section{Comparing DFT to QPI}

\begin{figure}[t]
    \includegraphics[width=\columnwidth]{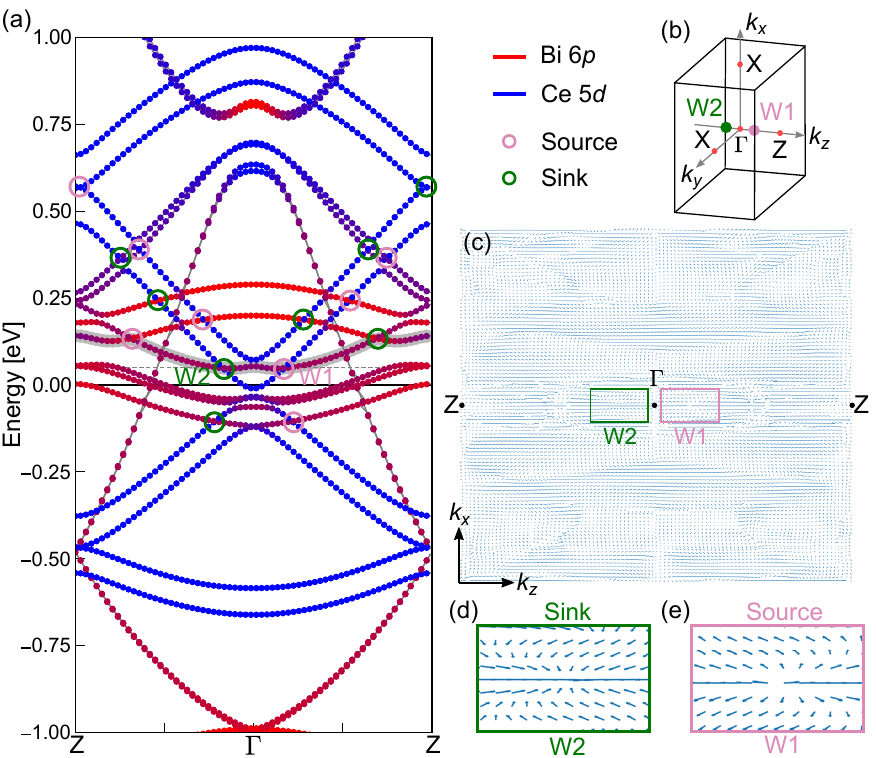}
    \caption{\textbf{Prediction of Weyl points in the \pppm\ phase of CeBi} (a) Band structure along the $\Gamma - Z$ direction in the ferrimagnetic \pppm\ phase, calculated using density functional theory (DFT). Open circles indicate the locations of Weyl nodes, colored to indicate sources (pink) and sinks (green) of Berry curvature.  (b) Three-dimensional Brillouin zone in the \pppm\ phase of CeBi, showing the Weyl nodes W1 and W2 closest to the Fermi level, around 50 meV [gray dashed line in (a)]. (c) Berry curvature field in $k_x - k_z$ plane of the gray highlighted band in (a), confirming the source (W1) and sink (W2) of Berry curvature. (d)-(e) Zoom on Berry curvature around W1 and W2.}
    \label{fig:Berry}
\end{figure}



Figure \ref{fig:Berry}(a) shows our DFT calculation of the band structure in the ferrimagnetic \pppm\ phase of CeBi, over a larger energy range than in Fig.\ \ref{fig:DFT}(d). To find the Weyl nodes, we calculate the Berry curvature at each crossing, and circle the sources (pink) and sinks (green). In Figs.\ \ref{fig:Berry}(c-e) we plot the Berry curvature of the Weyl points W1 and W2 closest to the Fermi level.  

\begin{figure}[t]
    \includegraphics[width=1\columnwidth]{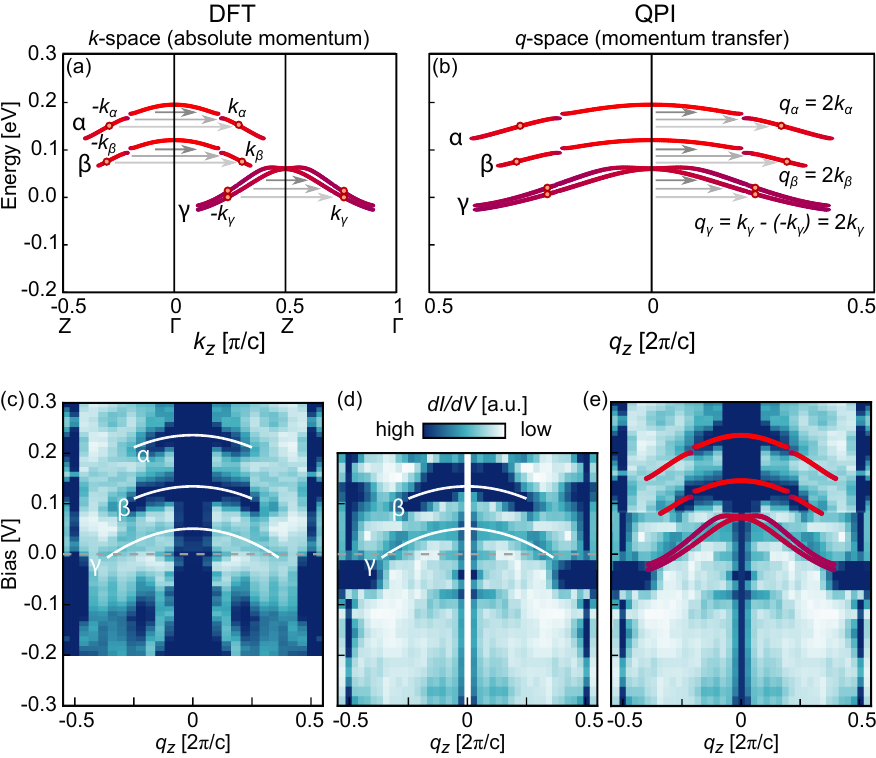}
    \caption{\textbf{Quasiparticle interference (QPI) in the \pppm\ phase at $B_z=3$ T.}
     (a-b) Relation between $k$-space and $q$-space in QPI measurements. (a) DFT-calculated segments of the Bi outer $p$ band, replicated from Fig.\ \ref{fig:DFT}(d), but with the $\gamma$ band shifted up by $\sim 70$ meV to match our measured data. Intra-band quasiparticles scatter from $-k$ to $+k$ with a total momentum transfer of $q = 2k$, shown as gray arrows. (b) The momentum transfer vectors ($q$) originate from zero at the tops of all three bands, $\alpha$, $\beta$, and $\gamma$. Therefore, pockets around $Z$ in $k$-space appear around zero momentum transfer in $q$-space.
    (c-e) Experimental QPI data and comparison to DFT. (c) QPI data set \#1 along the $\Gamma-Z$ direction, reproduced in Fig.\ \ref{fig:QPI}(a) for bias voltages above 80 mV. Setup parameters: $\Vs = 300$ mV, $\Is = 2$ nA, $\Vrms = 7.1$ mV.
    (d) QPI data set \#2 along the $\Gamma-Z$ direction, reproduced in Fig.\ \ref{fig:QPI}(a) for bias voltages below 80 mV. Setup parameters: $\Vs = 300$ mV, $\Is = 4$ nA, $\Vrms = 7.1$ mV.
    (e) Direct overlay of DFT-calculated band structure on combined QPI measurement. The bands with Bi $p$ orbital character, which define the intra-band scattering vectors $\alpha$, $\beta$, and $\gamma$, are consistent with the observed QPI intensity. Here the calculated $\gamma$ bands are shifted up by $\sim 70$ meV with respect to the calculated $\alpha$ and $\beta$ bands from Fig.\ \ref{fig:DFT}(d), indicating the presence of electron correlations.  
    }
    \label{fig:QPI_app}
\end{figure}

We investigate the band structure experimentally by imaging quasiparticle interference (QPI), which probes elastic momentum transfer, predominantly originating from intra-band scattering, as shown in Fig.\ \ref{fig:QPI_app}(a). Therefore, all scattering vectors appear around $q = 0$ in a QPI measurement, as shown in Fig.\ \ref{fig:QPI_app}(b). Figure \ref{fig:QPI_app}(c) and \ref{fig:QPI_app}(d) show energy vs.\ $q$ dispersion along the $q_z$ direction measured in two different energy ranges with similar setup conditions. The two datasets are combined in Figs.\ \ref{fig:QPI}(a) and \ref{fig:QPI_app}(e), where we overlay the calculated band dispersion ($\alpha$, $\beta$, $\gamma$) from the Bi $p$ orbitals. After a slight upshift of the $\gamma$ band, presumably caused by electron correlations, we find a good agreement between our DFT calculations and our QPI measurements. 

\clearpage

\section{Zero padding}
\begin{figure}[h]
    \includegraphics[width=1\columnwidth]{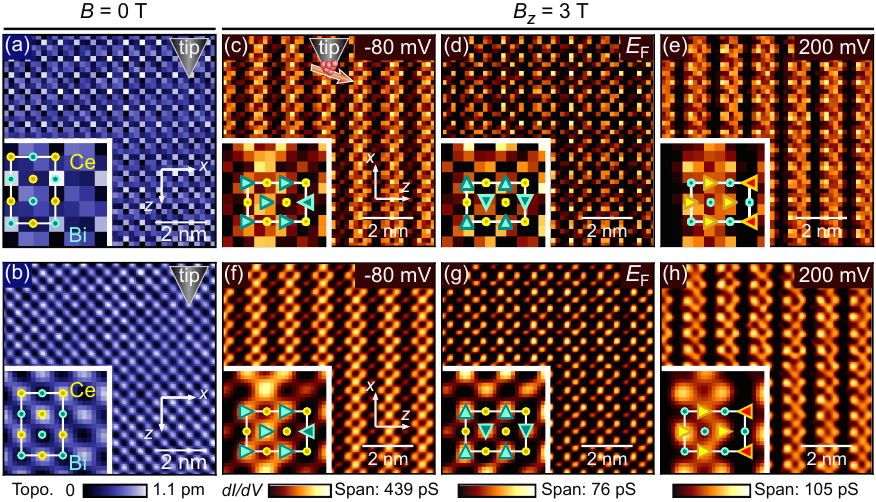}
    \caption{\textbf{Interpolation by zero-padding of the Fourier transform }
    (a-b) Raw (a) and interpolated (b) topography of Fig.\ \ref{fig:topos}(a).  (c-h) Raw (c-e) and interpolated (f-h) conductance maps of Figs.\ \ref{fig:dIdV}(a)-(c).  }
    \label{fig:interpol}
\end{figure}

In order to improve atomic visibility, the following maps were interpolated by Fourier-transforming, zero-padding the FT, then inverting the FT: Fig.\ \ref{fig:topos}(a), Fig.\ \ref{fig:dIdV}(a)-(c), Fig.\ \ref{fig:LaDoping}(f)-(j), Fig.\ \ref{fig:allenergies}(a), Fig.\ \ref{fig:canting}(a), Fig.\ \ref{fig:kondo_linecut}(a), and Fig.\ \ref{fig:kondo_linecut}(d). The comparison between raw and interpolated data is shown in Fig.\  \ref{fig:interpol}. 

\clearpage

%

\end{document}